# Recrystallization and Interdiffusion Processes in Laser-Annealed Strain-Relaxed Metastable Ge$_{0.89}$Sn$_{0.11}$

S. Abdi,[1] S. Assali,[1] M. R.M. Atalla,[1] S. Koelling,[1] J. M. Warrender,[2] and O. Moutanabbir[1,*]

[1] *Department of Engineering Physics, École Polytechnique de Montréal, C. P. 6079, Succ. Centre-Ville, Montréal, Québec H3C 3A7, Canada*

[2] *US Army Combat Capabilities Development Command – Armament Center, Benet Laboratories Directorate, Watervliet, New York 12189, USA*

**ABSTRACT:**

The prospect of GeSn semiconductors for silicon-integrated infrared optoelectronics brings new challenges related to the metastability of this class of materials. As a matter of fact, maintaining a reduced thermal budget throughout all processing steps of GeSn devices is essential to avoid possible material degradation. This constraint is exacerbated by the need for higher Sn contents along with an enhanced strain relaxation to achieve efficient mid-infrared devices. Herein, as a low thermal budget solution for post-epitaxy processing, we elucidate the effects of laser thermal annealing (LTA) on strain-relaxed Ge$_{0.89}$Sn$_{0.11}$ layers and Ni-Ge$_{0.89}$Sn$_{0.11}$ contacts. Key diffusion and recrystallization processes are proposed and discussed in the light of systematic microstructural studies. LTA treatment at a fluence of 0.40 J/cm$^2$ results in a 200-300 nm-thick layer where Sn atoms segregate toward the surface and in the formation of Sn-rich columnar structures in the LTA-affected region. These structures are reminiscent to those observed in the dislocation-assisted pipe-diffusion mechanism, while the buried GeSn layers remain intact. Moreover, by tailoring the LTA fluence, the contact resistance can be reduced without triggering phase separation across the whole GeSn multi-layer stacking. Indeed, a one order of magnitude decrease in the Ni-based specific contact resistance was obtained at the highest LTA fluence, thus confirming the potential of this method for the functionalization of direct bandgap GeSn materials.

## I. INTRODUCTION

GeSn group IV semiconductors have attracted high interest owing to their bandgap tunability in the short-wave infrared (SWIR) and mid-wave infrared (MWIR) spectral ranges, and their compatibility with the well-established Si complementary metal-oxide-semiconductor (CMOS) processing.[1–5] During the last decade, the significant progress in non-equilibrium growth processes has led to the development of device-quality GeSn heterostructures with Sn content exceeding by one order of magnitude its equilibrium solubility in Ge.[1] Typically, the growth is performed at temperatures lower than 400 ºC to avoid segregation of Sn that would degrade the material properties. However, this draws an upper limit in terms of the thermal budget needed in the subsequent steps for device processing and functionalization. This inherent limit in the allowed thermal budget must be considered in optimizing post-growth device fabrication such as the development of ohmic contacts, which are critical in electronic and optoelectronic devices. These applications require a low specific contact resistivity, $\rho_c$, on both n-type and p-type doped junctions. Achieving such performance could be especially challenging for n-type GeSn due to the Fermi-level-pinning (FLP) near the valence band edge, similar to Ge.[6] To circumvent this issue, semi-metallic nickel alloyed contacts NiGe and Ni(Ge$_{1-x}$Sn$_x$) have been proposed to act as an intermediary layer between the pure metal and Ge or GeSn contacted region, thereby reducing the FLP.[7] This process was explored earlier for GeSn at Sn-contents below 6 at.% using rapid thermal annealing (RTA) at temperatures below 400 ºC.[8,9] However, the use of RTA cannot be effective as the Sn-content increases due to the associated reduced thermal stability of GeSn, and the concomitant higher temperature needed for Ni(Ge$_{1-x}$Sn$_x$) phase formation. For instance, it was found that Ni(Ge$_{0.9}$Sn$_{0.1}$) forms at 80 °C higher than NiGe.[10] Moreover, RTA processing of GeSn with Sn-contents above 7-8 at.% at 460 ºC leads to severe phase separation and the formation of holes on the surface. Thus, RTA was found incompatible with the processing of Ni contacts on GeSn at 10 at.% Sn and above.[11]

The challenges above raise the need for different processes to meet the thermal processing requirements for $Ge_{1-x}Sn_x$. In this regard, laser thermal annealing (LTA) appears to be a promising method. It provides total control of the treated surface's depth and temperature and allows for ultra-fast heat dissipation at the microsecond scale, thus only affecting the surface. As such, LTA is suitable for shallow and fast thermal surface treatments required to process stacked epilayers such as multi-quantum wells, without promoting dopant diffusion or compromising the sharpness of their buried interfaces.[12,13] The underlying mechanisms and the limitations of this approach can be relatively easily controlled.[12–14] Moreover, LTA has already been successfully implemented in surface-related processes for a wide variety of material systems, such as improved electrical properties of hafnium oxide gate stacks.[15] In addition, LTA processing of metal-Ge contacts yields very high and shallow dopant activation well above $10^{20}$ cm$^{-3}$, exceeding their solid solubility limits in Ge.[16,17] Moreover, these contacts can be further enhanced, by forming continuous and atomically flat epitaxial nickel germanide layers ($NiGe_2$), which further reduces the FLP.[18,19] Consequently, the combination of high dopant activation near the surface, and higher quality of the formed germanide, resulted in Ni/n-Ge contacts with $\rho_c$ values that are 100× lower than that established by conventional RTA processing.[20] LTA is also well-suited to minimize the impact of thermal treatments on the structural properties of Ge and GeSn. This results from their high absorption coefficient at UV wavelengths and low melting temperatures.[12]

Note that recent studies on LTA-processed pseudomorphic GeSn layer showed the potential of pursuing this method to functionalize GeSn-based devices, by proving the existence of a large thermal budget range in which these devices can be processed without compromising the thin films' quality.[21,22] Notwithstanding these early studies, detailed investigations of the effects of LTA of the stability at higher Sn content are still missing despite the critical information they could provide on post-growth processing of direct bandgap GeSn materials. With this perspective, here we address the structural stability of LTA-processed metastable $Ge_{0.89}Sn_{0.11}$ multi-layers. In as-grown $Ge_{0.89}Sn_{0.11}$, LTA induces a defective layer with a thickness of 190 nm, where Sn-rich lines (15-20 at.%) extending all the way to the surface are visible within the $Ge_{0.89}Sn_{0.11}$ matrix. The same LTA treatment performed on Ni-$Ge_{0.89}Sn_{0.11}$ contacts reveal enhanced Ni diffusion along the Sn-rich dislocation lines in a similar fashion to pipe diffusion.[23] Moreover, V-shaped features develop when these segregation lines reach the Ni-GeSn interface and are possibly accompanied by the presence of voids. The LTA treatments results in a one order of magnitude decrease in the contact sheet resistance, which is possibly limited by local inhomogeneities at the Ni-GeSn/GeSn interface.

## II. EXPERIMENTAL DETAILS

The GeSn multi-layer structures were grown on 100 mm Si (100) wafers using a low-pressure chemical vapor deposition (CVD) system. First, a 1.6 μm-thick Ge layer, commonly refereed to as virtual substrate (Ge-VS), was grown on Si following a two-temperature step process (450/600 °C) and using post-growth thermal cyclic annealing at temperatures around 800 °C. The GeSn layers were grown at a pressure of 50 Torr, a constant $H_2$ flow and $GeH_4$ molar fraction of $1.2\times10^{-2}$, and the composition was controlled by the temperature change.[24,25] The four GeSn layers with increasing Sn content (schematics in Fig. 1a) were grown at 340 °C (#1), 330 °C (#2), 320°C (#3), and 310 °C (#4) by introducing SnCl$_4$ precursor. In addition, the initial SnCl$_4$ molar fraction (9.1×10$^{-6}$) was reduced by ~20 % during each temperature step to compensate for the reduced GeH$_4$ decomposition as the growth temperature drops.

Metal contacts were deposited using optical lithography in a transfer length method (TLM) configuration. Rectangular contacts with dimensions of 20 × 60 $\mu$m² and 30 × 60 $\mu$m² were patterned with spatial separations varying from 4.6 to 21.6 $\mu$m. The sample was cleaned in 1 % HF solution for 1 min prior to deposition of a 30 nm-thick Ni layer. LTA treatments were performed using frequency-tripled Nd:YAG laser (Ekspla) at a wavelength of 355 nm, with a fixed FWHM pulse duration of 6 ns. The 8 mm diameter laser beam was spatially homogenized and passed through a 3×3 mm² square aperture, such that a variation of the laser intensity was ensured to be lower than 10-15 %. One laser shot was delivered in each area before moving to the next location in raster mode. This covered a 2×2 grid with a minimum overlap of less than 10 μm between the grid cells. The shot-to-shot variation was ensured to be low. Four different energy densities of 0.20 J/cm², 0.25 J/cm², 0.30 J/cm², and 0.40 J/cm² were selected for the LTA experiments to span the various laser-GeSn interaction modes, which will be discussed later. All the energy fluences were calibrated by melting bare Si and Ge. Time-resolved reflectivity with a 496 nm laser was used to monitor the melt duration, which was then compared to one-dimensional heat flow calculations that employed conventional thermophysical parameters for Si and Ge. The treated samples were cleaned for several minutes in acetone before carrying out any further experimental characterization.

## III. RESULTS AND DISCUSSION

### A. Structural properties after LTA

X-ray diffraction (XRD) measurements were performed to investigate the effect of the LTA treatment on the crystalline quality and composition of the GeSn heterostructure. (2θ-ω) scans around the (004) XRD order are shown in Fig. 1b. The Ge-VS peak is visible at 66.06°, while the GeSn multilayer peaks are observed at lower angles. No measurable changes throughout LTA are detected in the XRD signal of the first three buffer layers (#1-3) in the range 64.7-65.5 °, while a strong reduction in the intensity of the $Ge_{0.89}Sn_{0.11}$ (#4) peak at 64.6° is visible as the LTA fluence increases. At the highest fluence of 0.40 J/cm$^2$, an additional diffraction peak (#4-LTA) is detected at ~65.2°. To decouple strain and composition effects, reciprocal space map (RSM) measurements around the asymmetrical (224) XRD peak were acquired on samples immediately after growth (Fig. 1c) and after LTA at 0.40 J/cm$^2$ (Fig. 1d). Because of the difference in the thermal expansion coefficient between Ge and Si, a residual tensile strain $\varepsilon_{||} < 0.2$ % is estimated for the Ge-VS. The peaks of four GeSn layers corresponding to Sn content of 4.5 at.% (#1), 5.5 at.% (#2), 7.8-9.1 at.% (#3), and 11.2 (#4) at.% are visible in the as-grown sample, with a compressive strain $\varepsilon_{||}$ below $-0.3$ %. After LTA at 0.40 J/cm$^2$, the $Ge_{0.89}Sn_{0.11}$ peak intensity decreases, and an additional signal associated with a tensile-strained $Ge_{0.91}Sn_{0.09}$ layer (#4-LTA) develops (dashed circle in Fig. 1d), in agreement with the (004) XRD scan in Fig. 1b. We note that no equilibrium phase (~1 at.% Sn) is detected, which suggests that significant Sn segregation and phase separation did not occur after LTA.

To investigate the evolution of the $Ge_{0.89}Sn_{0.11}$ surface, atomic force microscopy (AFM) measurements were performed. In the $10\times10$ $\mu m^2$ AFM map of the as-grown $Ge_{0.89}Sn_{0.11}$ sample in Fig. 2a, the typical cross-hatch pattern is observed, with a root mean square (RMS) roughness of ~14 nm. Similar morphology and surface roughness are recorded after annealing at 0.40 $J/cm^2$ (Fig. 2b-c). No macroscopic β-Sn phase droplets (typically with size of few micrometers) are visible on the surface.[23,26] However, by performing high-resolution $1\times1$ $\mu m^2$ AFM maps additional features can be observed after LTA treatment. While in the as-grown $Ge_{0.89}Sn_{0.11}$ sample a rather flat surface is visible (Fig. 2d), after LTA nano-islands with lateral dimensions lower than 100 nm are observed in the 0.40 $J/cm^2$ AFM map (Fig. 2d). The RMS roughness extracted from $1\times1$ $\mu m^2$ AFM maps was found to increase from 1 to 5 nm as LTA fluence increases (Fig. 2f).

The $Ge_{0.89}Sn_{0.11}$ microstructure after LTA at 0.40 $J/cm^2$ is highlighted in the cross-sectional transmission electron microscopy (TEM) images displayed in Fig. 3. In the GeSn multi-layer structure, dislocations are commonly observed in the lower Sn content buffer layers, while a high crystalline quality is generally obtained in the upmost region of the stacking.[24,25] Because of the LTA treatment, a highly defective (yet crystalline) layer with a thickness of ~190 nm is visible within the 260 nm-thick $Ge_{0.89}Sn_{0.11}$ layer (#4) (Fig. 3b). Higher resolution TEM (HRTEM) images in Fig. 3c-d reveal the presence of dislocations within a single crystalline layer, which terminate with a protruded surface. Interestingly, the dislocation lines do not extend deeper, but instead terminate about 50 nm above the depth reached by the laser. The composition of the LTA-affected area is estimated using energy dispersive X-ray (EDX) analysis in Fig. 4. The cross-sectional scanning TEM (STEM) image and EDX map for the Sn atoms are shown in Fig. 4a-b, together with the EDX line-scan across the whole multi-layer structure (Fig. 4e). The as-grown GeSn layers and LTA-altered region can be clearly identified in the STEM-EDX images. No visible changes are detected in the #1-3 layer, whereas a sharp drop in Sn content, that is followed by locally enriched regions, is visible in the #4 layer. Higher resolution STEM-EDX images (Fig. 4c-d) clearly show the presence of Sn-enriched vertical columns that extend from the surface down to ~25 nm before the Sn-depleted region, which is associated with the largest depth reached by the laser. We highlight that only limited surface Sn segregation is detected, both on the Sn-rich lines and on the regions with uniform composition, which differs from earlier observations on pseudomorphic $Ge_{0.83}Sn_{0.17}$ samples grown by molecular beam epitaxy.[21] From the kinetics of impurity trapping during the LTA process,[14,27] one would expect a high Sn concentration in the Sn-rich columns near the surface . However, this is not observed in our samples. To further show this, EDX line-scans recorded along a Sn-rich column and across the lateral direction are plotted in Fig. 4f and Fig. 4g, respectively. A monotonic increase in Sn incorporation from ~10 at.% to ~18 at.% (surface) is visible along the Sn-rich column, without a large accumulation of Sn at the surface. An average Sn content of ~11 at.% is estimated along the lateral direction, which is close to the 9-11 at.% range obtained from the RSM map (Fig. 1c-d). Considering the ~2 at.% inaccuracy in the EDX method, the measured value agrees well with the 9.3 at.% peak that develops in RSM after LTA (Fig. 1d). The origin of these defective, Sn-rich lines is still debated in literature. E. Galluccio et. al.[22] associated this behavior with the onset of cellular breakdown,[28,29] while T.T. Tran et. al. ruled out this mechanisms.[30] A more detailed discussion of the possible processes is provided in section C below.

## B. LTA on Ni/GeSn contacts

Like the as-grown samples, we also investigated the properties of LTA-annealed Ni-deposited Ge$_{0.89}$Sn$_{0.11}$. The cross-sectional TEM and STEM-EDX of GeSn with Ni top contacts after LTA at 0.40 J/cm$^2$ are displayed in Fig. 5 and Fig. 6, respectively. While this structure shares similarities with the case of pristine Ge$_{0.89}$Sn$_{0.11}$ (Fig. 4), without a large accumulation of Sn at the surface, the presence of Ni results in a richer morphological structure. First, the defective LTA-GeSn layer now extends up to a depth of 360 nm (Fig. 5a), hence beyond the initial 260 nm-thick Ge$_{0.89}$Sn$_{0.11}$ layer (#4) and covering the first 100 nm of the Ge$_{0.91}$Sn$_{0.09}$ layer (#3). Second, a NiGeSn layer forms at the surface with a rough, faceted NiGeSn/GeSn interface (Fig. 5b). Moreover, V-shaped NiGeSn structures are visible (Fig. 5b, 6c), which propagate all the way into the GeSn layer underneath as vertical columns. The presence of Moiré fringes within these filaments (Fig. 5c) indicates a misorientation between the lattice planes of the column *w.r.t.* the surrounding single-crystalline GeSn layer. No significant bulk diffusion of Ni into GeSn is detected in between the NiGeSn columns (Fig. 6c), which suggests their role in strongly enhancing Ni diffusion. This mechanism resembles the so called pipe diffusion, where the diffusivity is enhanced by few orders of magnitude in the neighboring region of a dislocation core.[31–34] The role of pipe diffusion on the stability of GeSn during strain relaxation and thermal annealing was recently highlighted,[23,34] and it could also be critical during LTA-recrystallization of the NiGeSn layers. Additionally, the V-shape of the NiGeSn features (Fig. 6c) closely matches the morphology observed in superlattices structures that are subjected to pipe diffusion.[32] Interestingly, voids seem to form on the V-shaped NiGeSn features (dark areas in Fig. 6a-b), possibly resulting from the rapid diffusion of Ni into the Sn-rich vertical filaments underneath during the solidification process.

The EDX line-scan estimated along a Sn-rich column and perpendicular to it are plotted in Fig. 7a-c. The thickness of the deposited Ni layer reduces from 30 nm to ~10 nm after LTA treatment because of the diffusion[35] of Ni atoms into GeSn. In this epilayer, a ~20 nm-thick $Ni_{0.7}Ge_{0.3}$ layer develops (with a negligible Sn content, below EDX quantification limit), followed by a $Ni_{0.35}Ge_{0.50}Sn_{0.15}$ layer at larger depth. Moving deeper into the heterostructure, the voids are encountered, followed by the Sn-rich lines with a stoichiometry of 70-80 at.% Ge and 30-20 at.% Sn, while the remaining 10 at.% content is associated with Ni (challenging quantification due to background). This is further shown by the EDX line-scan acquired in the horizontal plan within the Ni-GeSn contact in Fig. 7d. The stochiometric equivalent of a $Ni(Ge_{1-x}Sn_x)_2$ phase (with x<10 at.%) was previously demonstrated in Ge ($NiGe_2$) samples after LTA at 0.40 $J/cm^2$.[27] While a similar identification in our samples is challenging, as it would require a precise estimation of the TEM diffraction pattern, the EDX quantification might indicate the presence of a $Ni(Ge_{1-x}Sn_x)_2$ phase along the Sn-rich lines.

## C. Diffusion and intermixing processes

This section addresses the mechanisms that might be involved in the diffusion and segregation of atomic species in the Ni-GeSn heterostructure, namely pipe diffusion, spinodal decomposition, and cellular breakdown. During the LTA process the upmost region of the GeSn sample (≤360 nm in depth) is expected to highly expand in volume (or eventually melt) due to the large temperature gradient induced by the laser.[21,27] Upon cooling (and re-solidification), the lattice mismatch between the underlying GeSn layers and the LTA-treated GeSn is relieved by the nucleation of a high density of dislocations in the latter (Fig. 3 and 5a). The newly formed network of dislocations can serve as fast diffusion path for Sn atoms, since the diffusion coefficient around the dislocation core can be enhanced by few orders of magnitude compared to bulk diffusion.[31,32,36] This pipe diffusion process has been recently observed in GeSn.[23,34] In the present study, we found that dislocations are also enriched by ~5 at.% higher Sn content as compared to the surrounding GeSn matrix, regardless of the presence or not of a Ni layer on top. Hints for the enhanced mass transport through the dislocations are seen in the Ni atoms diffusing deep across the whole thickness of the LTA-processed GeSn layers, resulting in Ni- and Sn-rich columnar structures. The V-shape structures observed in the Ni-GeSn layers also point to the enhanced diffusion around dislocation cores. Another energetically favorable process is spinodal decomposition of GeSn during the cooling process.[37] The temperature ramp up and cool down during LTA certainly allows for the transition from the region between the binodal and spinodal curves to the region under the spinodal curve of GeSn at 11 at.% Sn.[37] Therefore, the timeframe for this transition may allow for phase separation in GeSn during LTA, although the resulting structures are usually interconnected and more entangled in the three dimensions.[38] One might consider the unidirectionality and ultra-fast quenching kinetics in LTA as conditions that would yield columnar structures instead. Hence, the columns here might be considered as the onset of the spinodal decomposition in GeSn.

In addition to the solute trapping of Sn in Ge that is expected in these ultra-fast solidification kinetics,[39] the 11× higher Sn content in our samples than equilibrium value (1 at.%) can cause cellular breakdown in the solidification front during laser annealing.[14,40] The combined effect of rapid solidification and lateral diffusion of Sn can cause a breakdown of the solidification front at a certain depth above the LTA-affected region limit.[40] This would lead to the formation of Sn-rich columns with a precise spacing that subsequently preserve their composition during solidification. Note that these regions would have a delayed solidification relative to the bulk as a result of the breakdown itself,[41] combined with the expected lower melting temperature of GeSn with higher Sn contents, given the difference in the melting temperatures of pure Ge and Sn of 938 °C and 232 °C, respectively.[42,43] Upon reaching the surface, these columns would normally translate into cellular structures.[41] However, such pattern was not observed in AFM images (Fig. 2). As for Ni enrichment in the columns, the reported solubility limit for Ni in liquid Ge is orders of magnitude higher than that for Ni in solid Ge,[27,44] which allows for Ni to dissolve much more rapidly in the former. The delayed solidification of these filaments might contribute to their Ni enrichment *w.r.t* the neighboring regions.

The presented data reveal the unlikelihood of the cellular breakdown mechanism at several levels. First, the critical solute concentration that would cause an unstable interface (and triggering cellular breakdown) depends on the solidification's front velocity.[40] No evident morphological change was observed in $Ge_{0.89}Sn_{0.11}$ samples treated at different LTA fluences. This would suggest that the Sn content used here may be well below the critical concentration. Indeed, similar Sn-rich filamentary structures were also observed at a significantly lower Sn content in $Ge_{0.94}Sn_{0.06}$.[30] Similar features were not apparent in other studies performed on pseudomorphic GeSn layers with Sn contents up to 17 at.% and treated with LTA under similar conditions,[21,42,43] which would further suggest the absence of a direct correlation with the critical solute concentration. Second, the columnar structures could translate into cell walls at the surface and create a cellular structure after breakdown,[28,29] as demonstrated in the breakdown of LTA-treated supersaturated Si alloys,[14,40] as well as in phase-field simulations that comprehensively tracked the solidification front's shape for this mechanism.[41] This pattern was not identified in our AFM images at any LTA exposure.

At the present stage, it is still challenging to provide a complete understanding of the diffusion and re-solidification processes that can take place during LTA of epitaxial metastable $Ge_{0.89}Sn_{0.11}$. Overall, the full understanding of such behavior can significantly help optimize the annealing conditions to lower the density of these laser-induced defects, and thereby reducing the laser-induced damage. Moreover, if cellular breakdown is the dominant mechanism, careful considerations of the LTA parameters need to be taken to suppress this breakdown. This case might require using slower treatments such as flash-lamp annealing (FLA), which was shown to suppress cellular breakdown of Si supersaturated with Ti,[14,45] and it was recently employed to fabricate $Ge_{0.955}Sn_{0.045}$ layers.[46]

## D. Electrical characterization

The effect of LTA on the contact resistance $\rho_C$ and sheet resistance $R_{SH}$ of Ni contacts were also investigated. The $\rho_C$ and $R_{SH}$ values for different LTA fluences estimated from transfer length measurements (TLM) are displayed in Fig. 8. In the as-grown $Ge_{0.89}Sn_{0.11}$, $\rho_C = (1.85 \pm 0.05) \cdot 10^{-4}\ \Omega \cdot cm^2$ and $R_{SH} = (805 \pm 15)\ \Omega \cdot cm^2$ are measured, hence consistent with previously estimated values for GeSn samples at similar composition.[4] A decrease in $\rho_C$ is observed as the LTA fluence increases (Fig. 8a), with a significant drop by 5-10 times ($\rho_C = (1.5 - 5) \cdot 10^{-5}\ \Omega \cdot cm^2$) when the LTA fluence is increased above 0.25 J/cm². The decrease in $\rho_C$ is accompanied by a simultaneous increase of $R_{SH}$ (Fig. 8b), eventually reaching values that are ~3 times larger than the as-grown sample ($R_{SH} = (1500 - 2500)\ \Omega \cdot cm^2$). The remarkable reduction in $\rho_C$ with LTA indicates that the local inhomogeneities that form in the Ni-GeSn layer upon LTA (e.g., voids in Fig. 6) do not have a detrimental effect on the properties of the contact. However, these inhomogeneities could prevent a further reduction in $\rho_C$ (blocking current flow), thus the measured values should only be considered as an upper limit and further optimization is required to prevent void formation. The formation of abrupt NiGe/NiGe$_2$ phases in bulk Ge after LTA was previously correlated with the reduction in $\rho_C$ after LTA.[27] Our EDX analysis (Fig. 6c and Fig. 7) indicates a more complex stoichiometry and morphology of the NiGeSn contacts. This would further complicate evaluating the impact of the different phases on the total sheet resistance and specific contact resistance of the fabricated contacts, which requires a dedicated comprehensive study to further elucidate and optimize the LTA process and post-LTA processing. The latter should include, for instance, proper surface treatments to remove the piled-up impurities expelled during the solidification. This process can become even more complex in the presence of a metallic contact layer.

## IV. CONCLUSIONS

We investigated the effect of LTA on the structural properties of strain-relaxed, metastable Ge$_{0.89}$Sn$_{0.11}$ semiconductors and on the formation of Ni/GeSn contacts. By increasing the LTA fluence up to 0.40 J/cm$^2$, the diffusion of Sn atoms toward the surface is triggered thereby leading to the formation of Sn-rich columnar structures in the 200-300 nm-thick LTA-affected region, while buried GeSn layers are preserved. Our systematic structural analysis indicates pipe diffusion as the most likely process to occur during LTA treatment. Moreover, the specific contact resistance of Ni/GeSn contacts was lowered by an order of magnitude after LTA, while displaying a rich morphological structure resulting from the Ni diffusion along the Sn-rich columns. This demonstrates the potential of this method to thermally activate GeSn Ni-based ohmic contacts, albeit the revealed complexity of the stoichiometry and morphology of these contacts requires further investigations to optimize the process.

## ACKNOWLEDGMENTS

The authors thank J. Bouchard for the technical support with the CVD system and A. Kumar for the XRD measurements. O.M. acknowledges support from NSERC Canada (Discovery, SPG, and CRD Grants), Canada Research Chairs, Canada Foundation for Innovation, Mitacs, PRIMA Québec, and Defence Canada (Innovation for Defence Excellence and Security, IDEaS).

## AUTHOR INFORMATION

Corresponding Author:

*E-mail: oussama.moutanabbir@polymtl.ca

Notes: The authors declare no competing financial interest.

The following article has been submitted to Journal of Applied Physics. After it is published, it will be found at https://aip.scitation.org/journal/jap.

# FIGURES CAPTIONS

**Figure 1.** (a) Schematics of the $Ge_{0.89}Sn_{0.11}$ multi-layer heterostructure grown on a 4" Ge-VS/Si wafer. (b) 2θ-ω scans around the (004) XRD order as a function of the LTA fluence. (c-d) RSM around the asymmetrical (224) reflection for the as-grown (c) and 0.40 J/cm$^2$ (d) $Ge_{0.89}Sn_{0.11}$ samples.

**Figure 2.** (a-b) 10 × 10 μm AFM maps for the as-grown (a) and 0.40 J/cm$^2$ (b) $Ge_{0.89}Sn_{0.11}$ samples. (c) Plot of the 10 × 10 μm RMS as a function of the LTA fluence. (d-e) 1 × 1 μm AFM maps for the as-grown (d) and 0.40 J/cm$^2$ (e) $Ge_{0.89}Sn_{0.11}$ samples. (f) Plot of the 1 × 1 μm RMS as a function of the LTA fluence.

**Figure 3.** (a) TEM image of the 0.40 J/cm$^2$ $Ge_{0.89}Sn_{0.11}$ sample. (b-d) Higher resolution TEM images of (a) showing the formation of vertical defect lines in the LTA-affected region.

**Figure 1.** STEM images (a,c) and EDX compositional maps (b,d) of the 0.40 J/cm$^2$ $Ge_{0.89}Sn_{0.11}$ sample. (e-g) EDX compositional profiles of the entire GeSn stacking (e), along the vertical (f) and in plane directions of the LTA-affected region (g).

**Figure 5.** (a) TEM image of the 0.40 J/cm$^2$ $Ge_{0.89}Sn_{0.11}$ sample with the Ni contact. (b-c) Higher resolution TEM images of (a) showing the formation of vertical defect V-shaped regions.

**Figure 6.** STEM images (a,b) and EDX compositional maps (c) of the 0.40 J/cm$^2$ $Ge_{0.89}Sn_{0.11}$ sample with Ni contact.

**Figure 7.** EDX compositional profiles extracted from Fig. 6c along the vertical Sn-rich columns (a-b), in plane region of the LTA-affected $Ge_{0.89}Sn_{0.11}$ stacking (c), and in plane region in the proximity of voids (d).

**Figure 8.** Specific contact resistance $\rho_C$ (a) and sheet resistance $R_{SH}$ (b) as a function of the LTA fluence.

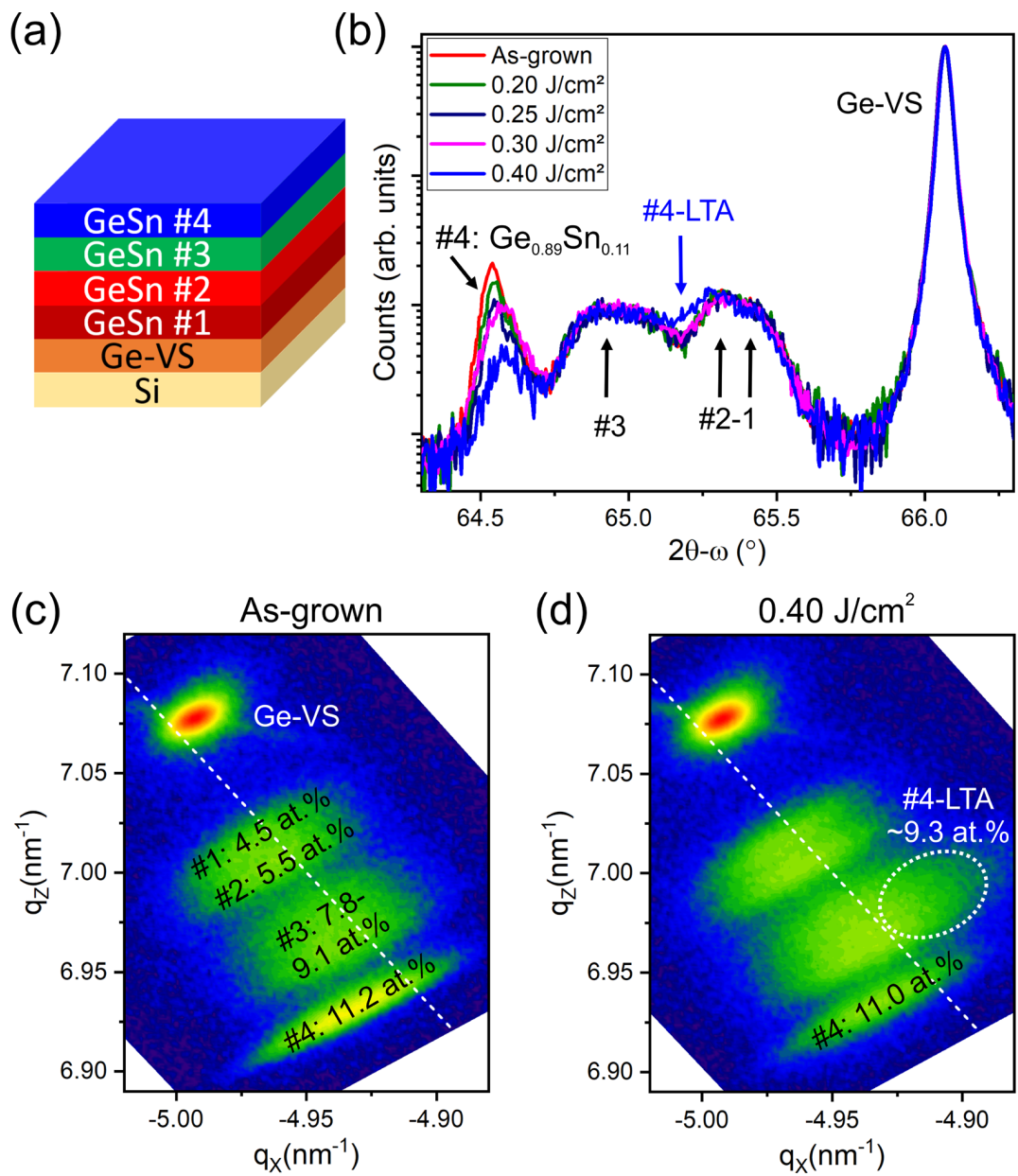

Figure 1

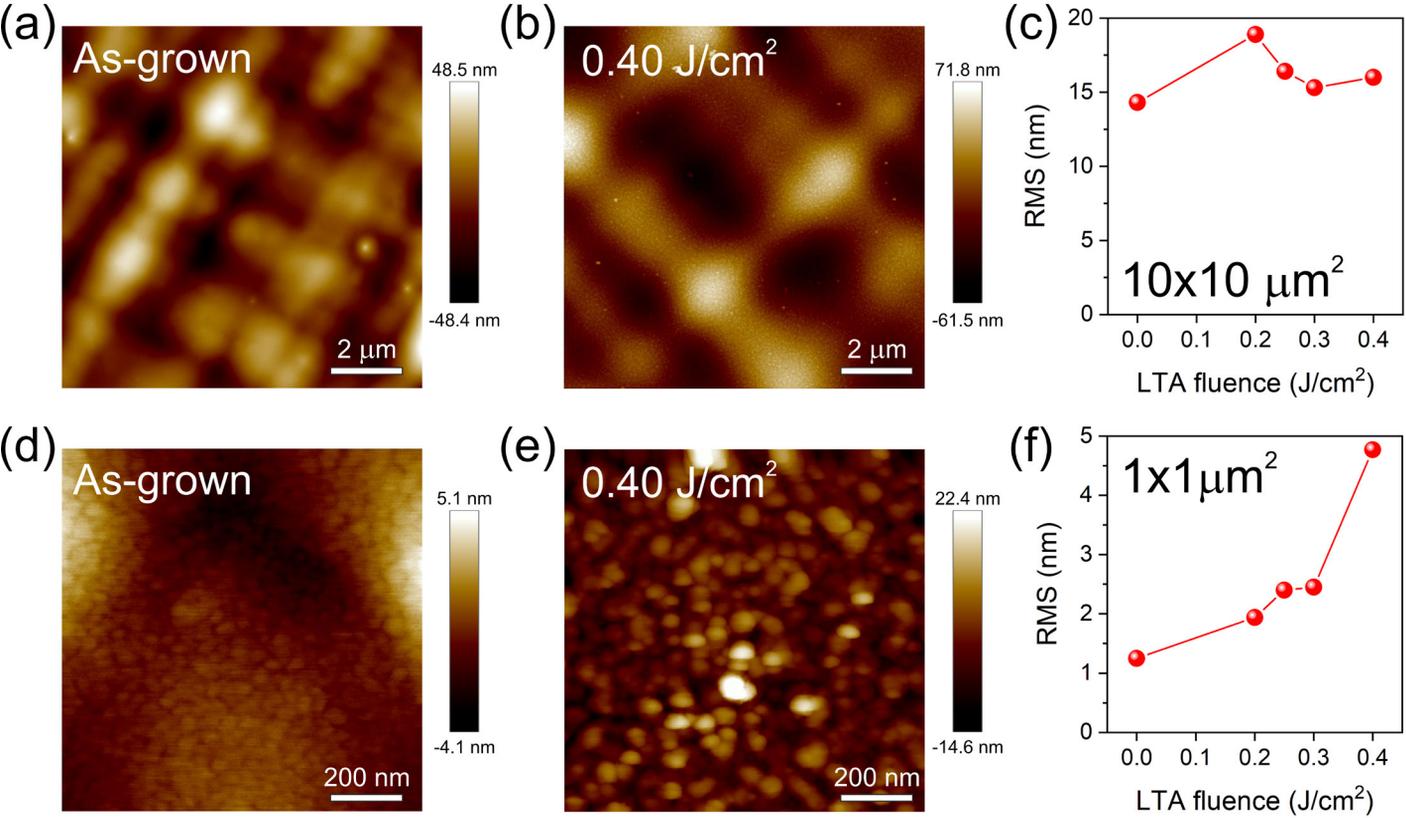

**Figure 2**

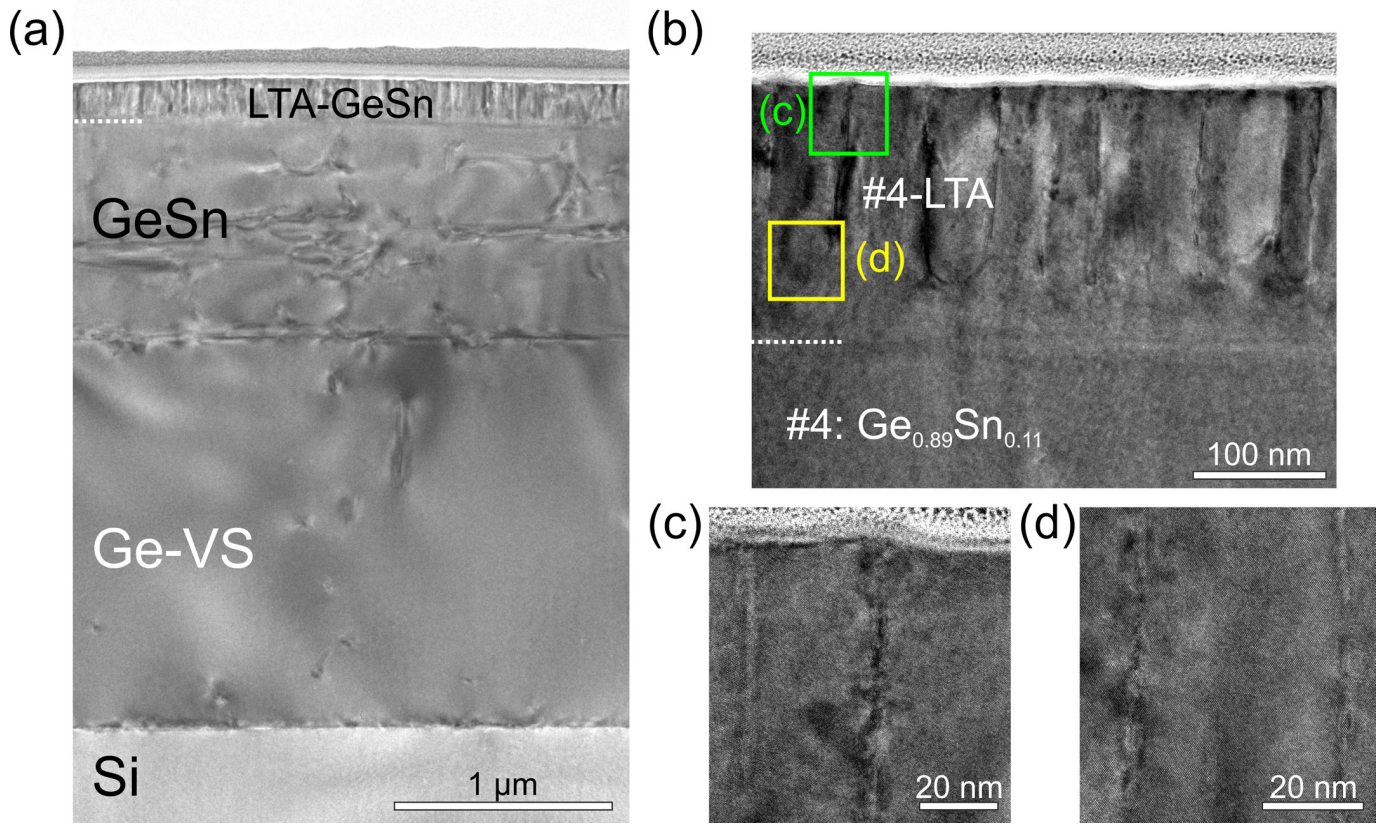

Figure 3

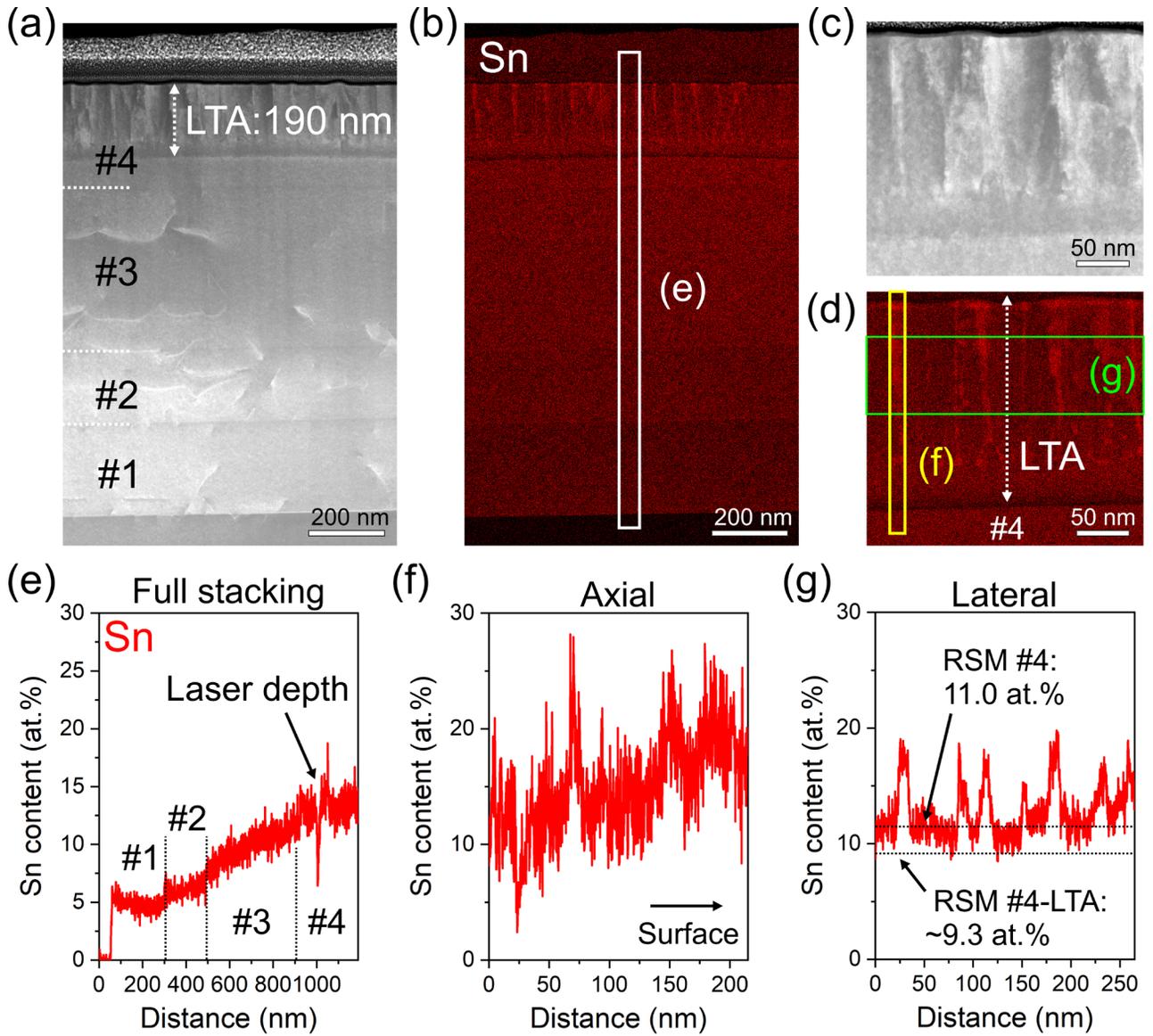

Figure 4

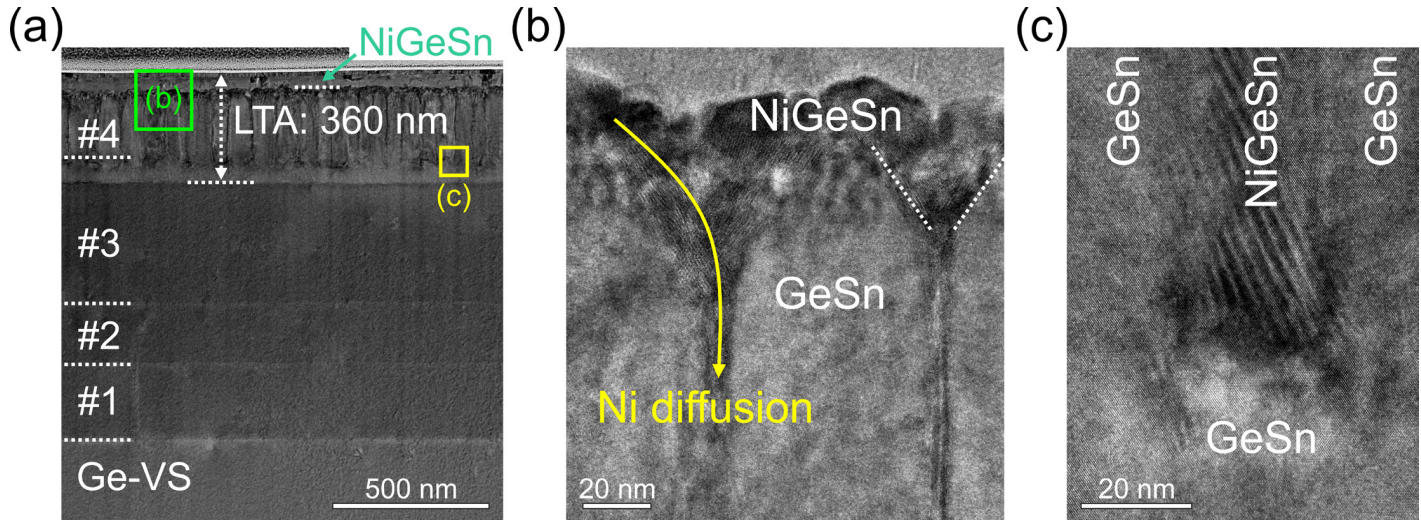

**Figure 5**

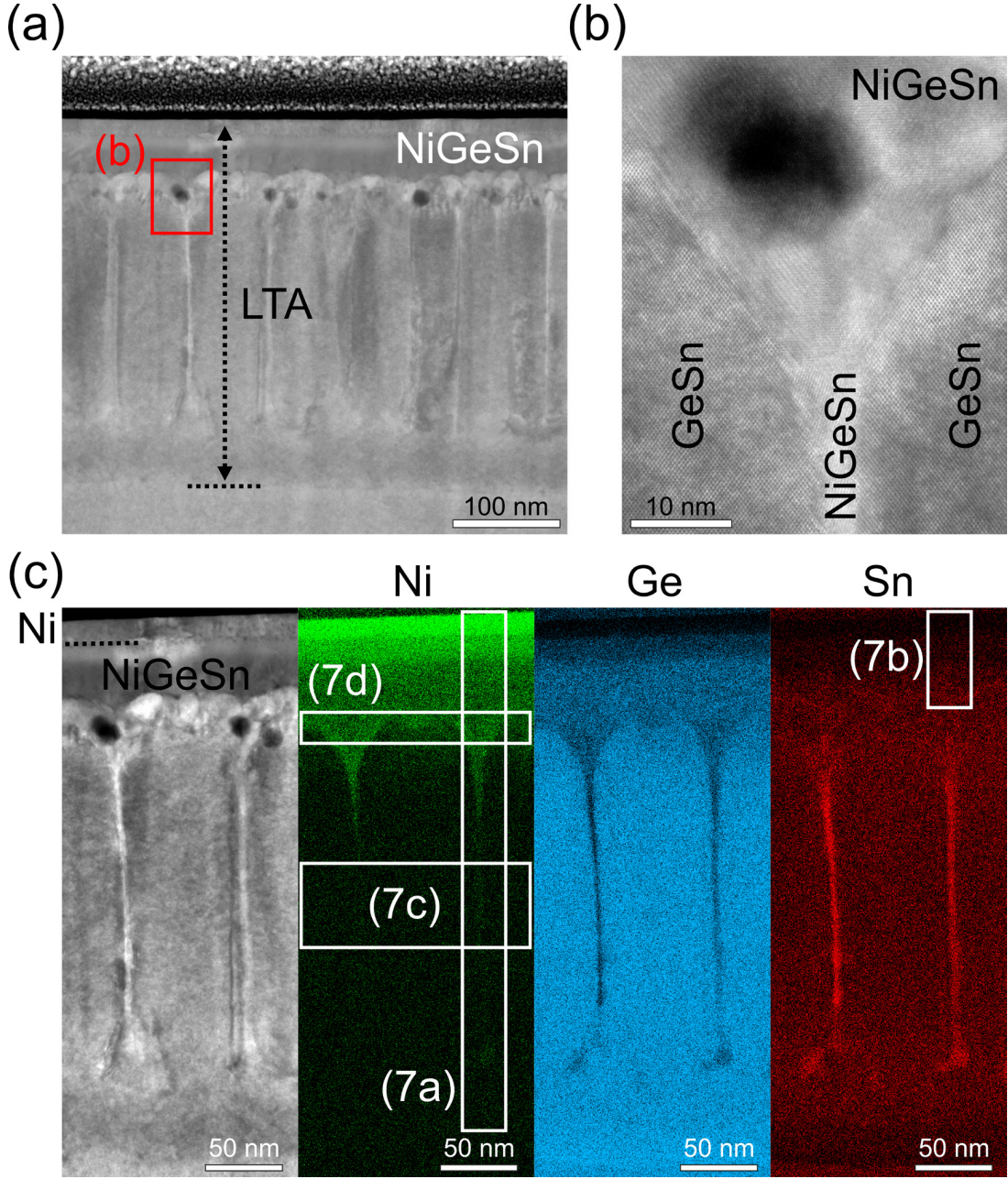

**Figure 6**

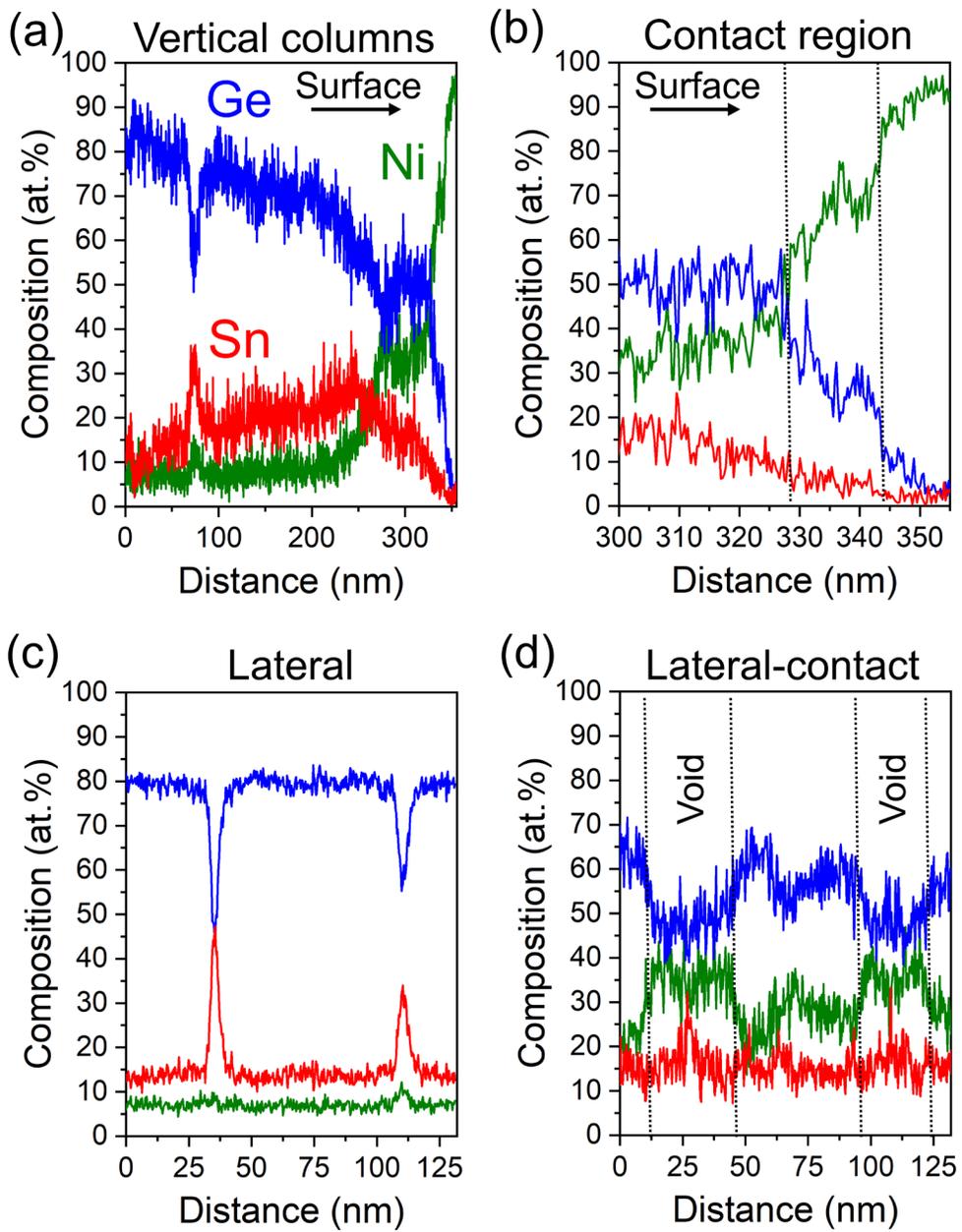

Figure 7

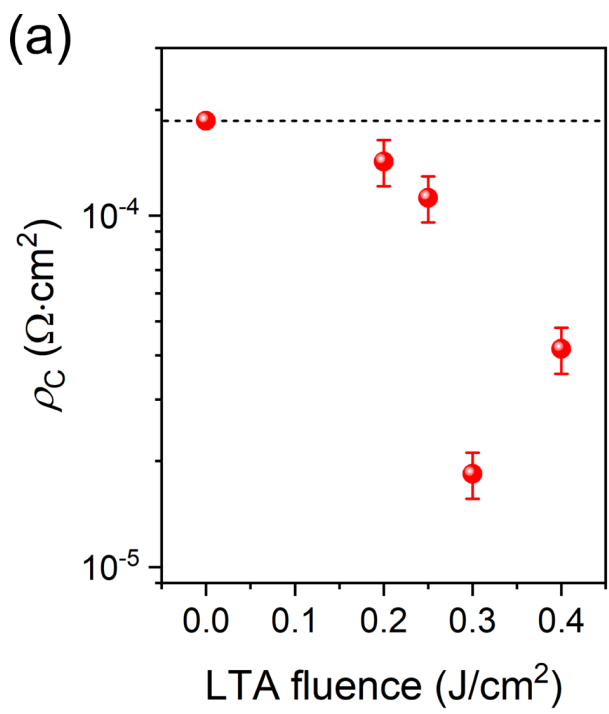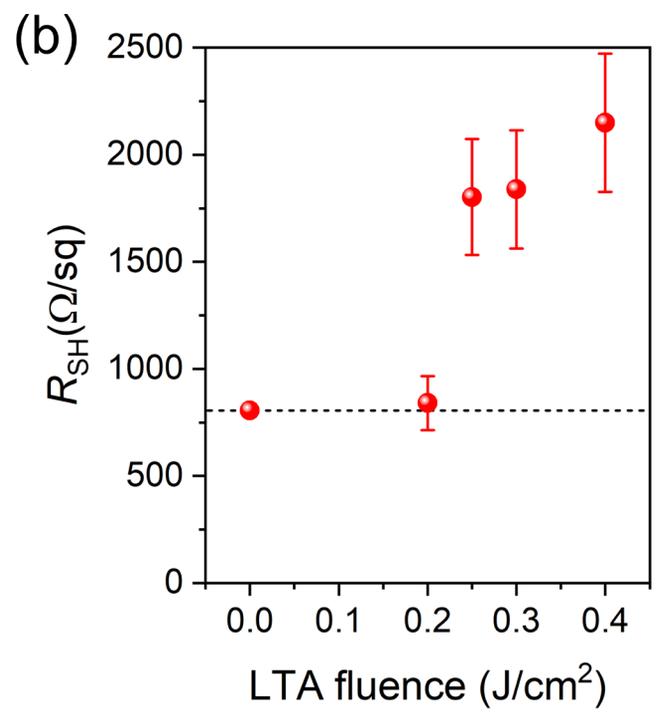

Figure 8